\documentstyle[aps,epsfig,multicol]{revtex}
\newcommand{\ri}{{\rm i}}
\newcommand{\re}{{\rm e}}

\newcommand{\rF}{{\rm F}}
\newcommand{\bq}{{\bf q}}
\newcommand{\tr}{{\rm tr}}
\newcommand{\ts}{\tilde{\sigma}}
\newcommand{\tL}{\tilde{L}}
\begin{document}

\title{Boundary Conditions, the Critical Conductance Distribution, and
One--Parameter Scaling}
\author{Daniel Braun$^{(1)}$, Etienne Hofstetter$^{(2)}$, Gilles
Montambaux$^{(3)}$, and Angus MacKinnon$^{(4)}$}
\address{$^{(1)}$ IBM T.S. Watson Research Center, P.O.~Box 218, Yorktown Heights, NY 10598\\
$^{(2)}$ Institute for Quantitative Finance, Imperial College, London SW7
2BX, England\\
$^{(3)}$ Laboratoire de Physique des
Solides,  associ\'e au CNRS \\
Universit\'{e} Paris--Sud, 91405 Orsay, France\\
$^{(4)}$ Blackett Laboratory, Imperial College, London SW7 2BW, England}

\maketitle
\widetext
\centerline{\today}
\begin{abstract}
\begin{center}
\parbox{14cm}{We study the influence of boundary conditions
transverse to the transport direction for disordered mesoscopic conductors
both at the Anderson metal--insulator transition and in the metallic
regime. We show that the boundary conditions  strongly influence the
conductance distribution exactly at the metal--insulator transition and we
discuss  implications for the standard picture of
one--parameter scaling. We show in particular that the scaling
function that describes the change of conductance with system size depends
on the boundary conditions from the metallic regime up to the
metal--insulator transition. An experiment is proposed that
might test the correctness of the one--parameter scaling theory.\\
PACS numbers: }
\end{center}
\end{abstract}

\begin{multicols}{2}
\section{Introduction}
More than fourty years after its discovery by Anderson \cite{Anderson58} the
disorder--induced metal--insulator transition is still the subject of strong
theoretical as well as experimental research
\cite{Kramer93}. One of the major achievements in the
long history of the Anderson metal--insulator transition (MIT) is the
renormalization group theory, which has also become known as
one--parameter scaling theory \cite{Abrahams79}. Its basic assumption is
that close to the transition the change of the dimensionless conductance
$g$ with the sample
size $L$ depends only on the conductance itself and not separately on energy,
disorder, the size of the sample, its shape, the elastic mean free path $l_\re$,
etc.. Many predictions,
like the lower
critical dimension, or the critical behavior \cite{Wegner85,Vollhardt82}
were successfully based on this theory, as well as an enormous amount of
numerical work that aimed at the direct calculation of the scaling function
$\beta(g)=d \ln g/ d\ln L$. Another important consequence of the one
parameter scaling theory is the prediction of a universal conductance
distribution $P^*(g)$ exactly at the MIT \cite{Shapiro86}. Earlier
numerical
work on the three dimensional Anderson model seemed to confirm the
universality of the conductance distribution \cite{Markos94}. The
dependence on the universality class was stressed in
\cite{Slevin97}.

Recently, however, some doubts have been cast on whether the
conductance distribution is universal within the {\em same}
universality class. Two different numerical studies reported two
different forms of $P^*(g)$ for the same system
\cite{Slevin97,Soukoulis99}, and it was found that the difference
originates in the use of different boundary conditions (BCs)
\cite{Slevin99}.

  The idea that $P^*(g)$ might depend on the BCs
appears indeed very natural after the discovery that spectral statistics, and
in particular the energy level spacing distribution $P(s)$ exactly at
the MIT, {\em do} depend on the BCs \cite{Braun98}. Samples
with periodic boundary conditions show a much stronger level repulsion than
samples with hard walls (Dirichlet boundary conditions).

In this work we show with a numerical analysis of the conductance
distribution at the critical point that $P^*(g)$ does indeed depend on the
BCs applied perpendicular to the transport direction. Choosing the
appropriate boundary conditions, we can reproduce both the results of
refs.\cite{Slevin97} and \cite{Soukoulis99}. In particular, the average
critical conductance $g_c$ depends on the BCs. This alone already implies a
dependence of $\beta(g)$ on the BCs since $g_c$ is defined as
$\beta(g_c)=0$. We confirm the BC--dependence of $\beta(g)$
analy\-ti\-cal\-ly by reinvestigating its form in the metallic regime with the
help of a $1/g$ expansion. Much to our surprise we find that earlier
analyses overlooked the effect of the BCs by approximating a sum over
diffusion modes by an integral. Evaluating the sum more carefully, we not
only find a dependence on the BCs, but also a so far unkonwn $\ln(l_\re/L)/g$
term in $\beta(g)$ in three dimensions that makes $\beta(g)$ non--universal
in the metallic regime.

\section{Numerical investigation at the Anderson transition}

The model studied is the three dimensional tight binding Anderson
Hamiltonian with diagonal disorder on a simple cubic lattice,
\begin{eqnarray} \label{HAnd1}
H&=&\sum_i e_i |i\rangle\langle  i|+u \sum_{\stackrel{<ij>}{\rm bulk}} |i\rangle\langle
j|\nonumber\\
&&+u\sum_{\stackrel{<ij>}{\sigma_y,\sigma_z}} c(
\re^{2\pi\ri\phi}|i\rangle\langle j|+h.c.)\,.
\end{eqnarray}
The $e_{i}$ are distributed uniformly and independently between
$-w/2$ and $w/2$. The notation $<ij>$ means next nearest neighbors,
$u$ is the hopping matrix element which we set equal to unity in
the following, and $w$ is the disorder parameter. The last sum in
eq.(\ref{HAnd1}) links corresponding sites on opposite sides of the
cubic sample perpendicular to the $y$ and $z$ directions, assuming
that transport occurs in the $x$--direction. Hopping between these
boundary sites arises when the system is closed to a ring ($c=1$)
and includes a phase factor $\re^{\ri 2\pi\phi}$, where $\phi$ is
the magnetic flux in units of $h/e$ inclosed by the ring. Hard wall
(Dirichlet) BCs correspond to $c=0$. The model (\ref{HAnd1}) shows
a MIT at the critical disorder $w_c\simeq 16.5$ \cite{MacKinnon83}.

The numerical calculation of the conductances uses a standard Green's
function recursion technique \cite{macKinnon85} that yields the
transmission matrix $t$ of the sample. The latter is connected to the
two--probe
conductance of the sample by the
Landauer--B\"uttiker formula
\begin{equation} \label{g3}
g=\tr\, tt^+\,,
\end{equation}
where $g=G/(e^2/h)$ denotes the conductance $G$ in
units of the inverse of the von Klitzing constant $h/e^2$.
Whether the two--probe conductance formula or the four--probe conductance
formula is used, is quite irrelevant at the metal insulator transition,
since the bulk resistance always dominates largely over the contact
resistance \cite{Braun97}. All conductances were calculated at energy
$E\simeq 0$. The number of conductances used for each BC and system size
ranged between $10^5$ for $L=6$ and $L=8$  to $2\cdot 10^3$ for $L=16$. All
system sizes $L$ are measured in units of the lattice constant.

\noindent
\begin{minipage}{3.38truein}
\begin{center}
\begin{figure}[h]
\epsfig{file=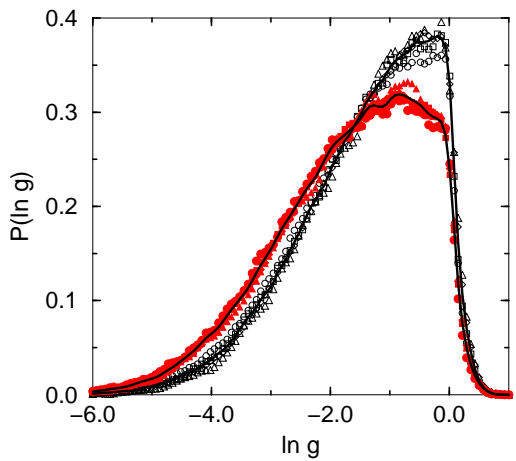,width=8cm}
\caption{ Critical conductance distribution for periodic and hard wall
boundary conditions, and different sample sizes. Sample sizes $L=8,10,12,16$
are denoted by circles, squares, diamonds and triangles, respectively; open
symbols indicate periodic BCs, full ones hard walls. The full lines are
averages over the above system sizes.
  \label{fig.Poflng}}
\end{figure}
\end{center}
\end{minipage}
\vspace*{0.3cm}

Our main numerical result is shown in Fig.\ref{fig.Poflng}, where we have
plotted the
distributions of the logarithm of the conductance at the transition for
periodic and  hard wall (HW) BCs, and different system sizes. For the same
BC the distribution is
almost independent of the system size, as is to be expected from the
criticality of the
ensemble at $w_c=16.5$. But the distributions are clearly very different for
the two BCs. The maximum of the distribution is considerably more pronounced
for periodic BCs than for hard walls. A more
detailed statistical analysis is presented
in table \ref{tab} and for the average values $\langle \ln g\rangle$  in
figure  \ref{fig.stat}.

\noindent
\begin{minipage}{3.38truein}
\begin{center}
\begin{table}[h]
\begin{center}
\begin{tabular}{c|c||c|c|c|c|}
L&BC&$\langle g\rangle=g_c$&$\sigma_g$&$\langle \ln g\rangle$&$\sigma_{\ln g}$\\\hline
6&P&0.356&0.314&-1.554&1.183\\
8&P&0.377&0.324&-1.476&1.159\\
10&P&0.392&0.329&-1.412&1.129\\
12&P&0.402&0.334&-1.378&1.118\\
16&P&0.413&0.336&-1.329&1.092\\\hline

6&HW&0.313&0.306&-1.777&1.281\\
8&HW&0.326&0.310&-1.710&1.252\\
10&HW&0.331&0.312&-1.685&1.246\\
12&HW&0.338&0.311&-1.675&1.211\\
16&HW&0.348&0.319&-1.614&1.222
\end{tabular}
\vspace*{0.2cm}
\caption{Statistical analysis of the critical conductance
distribution for
different boundary conditions ($P$ periodic and $HW$
hard wall). Besides the averages of $g$ and $\ln g$ also the standard
deviations
of these quantities, $\sigma_g$ and $\sigma_{\ln g}$ are given. \label{tab}}
\end{center}
\end{table}
\end{center}
\end{minipage}
\noindent
\begin{minipage}{3.38truein}
\begin{center}
\begin{figure}[h]
\epsfig{file=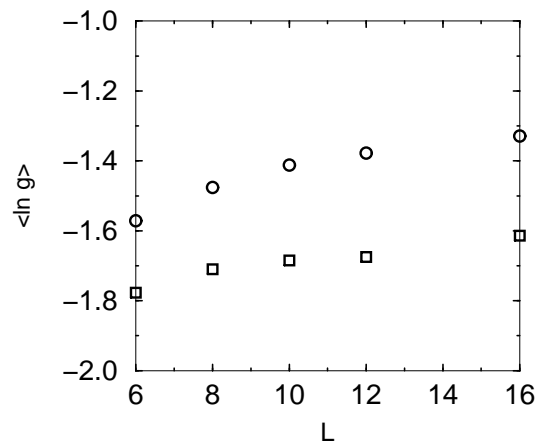,width=8cm}
\caption{As a function of system size the average $\langle \ln g\rangle$ is
  plotted for periodic (circles) and hard wall
  boundary conditions (squares).
  \label{fig.stat}}
\end{figure}
\end{center}
\end{minipage}
\vspace*{0.3cm}

The average is always over the disorder
ensemble. Fig.\ref{fig.stat}
shows that the average logarithmic conductance still depends slightly on the
system size in the regime investigated. But the difference between periodic
and hard wall BCs  does not diminish with increasing $L$, and
the dependence on $L$ decreases for larger $L$.
Where we have used the same system sizes as in
Refs.\cite{Slevin97,Soukoulis99} our values for all quantities calculated
($\langle g\rangle$,  $\langle \ln g\rangle$, and the standard deviations of
$g$ and $\ln g$) coincide within one percent with the values
given in these references. For comparison with \cite{Soukoulis99} $g$ should
be multiplied with a factor $2$, since we consider only one spin direction.
Thus, the discrepancy between \cite{Slevin97} and \cite{Soukoulis99} can
indeed be explained by the influence of the BCs (see also \cite{Slevin99}).\\

Our result has important implications for the scaling theory of the
metal--insulator transition, since it shows that the scaling
function $\beta(g)$ must depend on the BCs. The conductance that
enters in this equation has to be understood as an average
conductance \cite{Altshuler94}, and the critical conductance is
given by $\beta(g_c)=0$. According to our results $g_c$ depends on
the BCs, $g_c=0.413$ for periodic BCs and $g_c=0.348$ for hard
walls at $L=16$ (see table \ref{tab}), and therefore the $\beta(g)$
curves must at least be shifted as a function of the BCs. In the
next section we show by re--examining the weak localization
corrections to the conductance that also in the metallic regime
$\beta(g)$ depends on the BCs.

\section{Metallic regime}
It is well known that already in metallic regime $g\gg 1$ the quantum
interference
of diffusing electrons reduces the conductance compared to the classical
value $g=\sigma L$, where $\sigma$ is bulk conductivity.
The weak localization
correction $\delta g$ is given by a sum over diffusion modes as
\cite{Altshuler94}
\begin{equation} \label{dg}
\delta g=-2 \sum_{{\bf q}}\frac{\re^{-D\bq^2\tau_\re}}{\bq^2L^2}\,.
\end{equation}
 The sum  is limited  to the diffusive regime where $D\bq^2\ll
1/\tau_\re$. This limitation is taken into account by the exponential
cut-off; $\tau_\re$ is the elastic collision time, $D=v_\rF^2\tau_\re/3$ denotes the
diffusion coefficient and $v_\rF$ the Fermi velocity. The sum (\ref{dg})
depends on the BCs via the quantization
condition for the diffusion modes ${\bf q}$. For the transport
direction the wave vector is quantized according to $q_x=n_x\pi/L$, $n_x=1,2,\ldots$. Periodic boundary
conditions in the $y$-direction imply $q_y=n_y 2\pi/L$, $n_y=\pm 1,\pm
2,\ldots$ and correspondingly for the $z$-direction. Hard wall BCs on the
other hand lead to $q_y=n_y\pi/L$, $n_y=0,1,2,\ldots$ and $q_z=n_z\pi/L$,
$n_z=0,1,2,\ldots$. Consequently we have
\begin{equation} \label{dg1}
\delta g=-\frac{2}{\pi^2}S_{BC}(y)
\end{equation}
where the index $BC$ stands for a boundary condition and
\begin{eqnarray}
S_{P}(y)&=&\sum_{\stackrel{n_x>0}{n_y,n_z\ne
0}}\frac{\exp\left[-\pi^2(n_x^2+4n_y^2+4n_z^2)y
\right]}{n_x^2+4n_y^2+4n_z^2}\,,\\
S_{HW}(y)&=&\sum_{\stackrel{n_x>0}{n_y,n_z\ge 0}}\frac{\exp\left[-\pi^2(n_x^2+n_y^2+n_z^2)y \right]}{n_x^2+n_y^2+n_z^2}\,.
\end{eqnarray}
The argument $y$ is defined as
\begin{equation} \label{y}
y={D \tau_e \over L^2}= \frac{1}{3}\left(\frac{l_\re}{L}\right)^2\,.
\end{equation}
Previous analyses in the literature proceeded by approximating the
sum by an integral \cite{Altshuler94}, whereupon all dependence on the
boundary conditions is lost. While this is a good approximation for
$g\to\infty$, important corrections of the order $(\ln g )/g$ arise for
finite $g$, which we are going to derive now, assuming that to this order no
further diagrams beyond the diffuson approximation contribute. In
\cite{Kravtsov94}  it was shown by field theoretical methods combined with a
renormalization group approach that the diffuson approximation gives the
leading perturbative contribution to  the small
energy behavior of the spectral correlation function
to order $1/g^2$. \\
In order to proceed it is
convenient to differentiate $S_{BC}(y)$. The derivatives for both PBCs and HW
BCs can be written with the help of the function
\begin{equation} \label{F}
F(y)=\sum_{n=1}^\infty \re^{-\pi^2n^2 y}
\end{equation}
 as
\begin{eqnarray}
\partial_yS_{P}(y)&=&-\pi^2F(y)(2F(4y))^2\label{dySpbc}\,,\\
\partial_yS_{HW}(y)&=&-\pi^2F(y)(1+F(y))^2\label{dyShw}\,.
\end{eqnarray}
The function $F(y)$ is related to the complete elliptic integrals $K\equiv K(k)$ and
$K'\equiv K(k')$ with $k'=\sqrt{1-k^2}$ by \cite{RG}
\begin{equation} \label{FeI}
\frac{1}{2}\left(\left(\frac{2K}{\pi}\right)^{1/2}-1\right)=\sum_{n=1}^\infty
\re^{-\pi n^2 K'/K}\,.
\end{equation}
Since we are interested in $y\ll 1$, we need $K'/K\ll 1$ and therefore $k\to
1$ ($k'\ll 1$). For small values of $k'$ the elliptic integrals behave like
\begin{equation} \label{KK'}
K=\ln \frac{4}{k'}+{\cal O}(k'^2)\mbox{, }K'=\frac{\pi}{2}+{\cal O}(k'^2)\,,
\end{equation}
and we therefore obtain
\begin{equation} \label{Fy}
F(y)\simeq \frac{1}{2}\left(\left(\frac{1}{\pi y}\right)^{1/2}-1\right)\,.
\end{equation}
Inserting this into eqs.(\ref{dySpbc}), (\ref{dyShw}) and integrating with
respect to $y$ yields
\begin{eqnarray}
S_{P}&=&\frac{\sqrt{\pi}}{4\sqrt{y}}+\frac{5}{8}\pi\ln
y-2\pi^{3/2}\sqrt{y}+\frac{\pi^2}{2}y-\alpha_{P}\label{Spbc}\\
S_{HW}&=&\frac{\sqrt{\pi}}{4\sqrt{y}}-\frac{1}{8}\pi\ln y+\frac{1}{4}\pi^{3/2}\sqrt{y}+\frac{\pi^2}{8}y-\alpha_{HW}\label{Shw}\,.
\end{eqnarray}
whereas, replacing the sum (\ref{dg}) by an integral, one would have
found
\begin{equation}
S=\frac{\sqrt{\pi}}{4\sqrt{y}} -\alpha \,.
\end{equation}
where $\alpha$ is an integration constant resulting from the
cut-off at small $q \simeq 1/ L$. Thus, the leading term for small
$y$, $\sqrt{\pi/y}/4$, is the same for both boundary conditions.
The integration constants $\alpha_{P}$ and $\alpha_{HW}$ can be
evaluated numerically, by subtracting from the exactly calculated
sums the ana\-ly\-ti\-cal formulae (\ref{Spbc}) and (\ref{Shw})
without the constants. At the same time this serves as a sensitive
check for the correctness of these formulae. For small $y$ the
differences converge to
\begin{equation} \label{alphas}
\alpha_{P}\simeq -6.1509\mbox{, } \alpha_{HW}\simeq 2.3280\,.
\end{equation}
We have evaluated the sum numerically down to values $y=10^{-6}$, where in
particular the logarithmic term with the prefactors given above could be
clearly verified. \\
With Eqs.(\ref{Spbc}) and (\ref{Shw})
the conductance as a function of the dimensionless length $\tilde{L}\equiv
L/l_\re$ takes the form
\begin{equation} \label{gofL}
g=(\tilde{\sigma}-A)\tilde{L}-a \ln \tL+b+{\cal O}(1/\tL)
\end{equation}
for both periodic and hard wall BCs. The dimensionless bulk conductivity
$\tilde{\sigma}$ is defined as $\ts=\sigma l_\re h/e^2$, and the constant
$A=\sqrt{3}/(2\pi^{3/2})$ is the same for both BCs. The
coefficients $a$ and $b$ on the other hand do depend on the boundary
conditions; their values are
given in table \ref{tabc}.
Note that in the traditional approach the
coefficient $a$ vanishes!

Quite surprisingly $a<0$ for PBCs, which means
that the conductance increases even slightly faster than linearly with
the system size. This looks as if  there was anti--localization, but it should
be noted that the leading behavior due to weak localization is still the
usual decrease of the (bulk--)conductivity, i.e.~the leading term is
linear in the system size and with the expected negative sign.
The
fact that $a<0$  only for PBCs suggests a simple physical explanation for
the logarithmic term: Closing the
sample to a double torus by imposing PBCs allows for additional paths that
interfere
constructively and lead to enhanced localization for small system
sizes compared to the HW case. When increasing the system size these
additional localizing paths
quickly stop contributing and the conductance therefore increases more
rapidly than what would be expected just from the volume part of the weak
localization. \\

We are now in the position to explore the consequences of the BC dependent
weak localization corrections for the scaling function $\beta(g)$. Inserting
(\ref{gofL}) into the definition
\begin{equation} \label{bdef}
\beta(g)\equiv \frac{d \ln g}{d \ln \tL}
\end{equation}
yields
\begin{equation} \label{bL}
\beta(g)=1+\frac{1}{g}\left(a \ln\tL-b-a+{\cal
O}(1/\tL)\right)\,.
\end{equation}
It remains to reexpress $\tL$ by $g$. To this end we invert $g(\tL)$ from
(\ref{gofL}) to order $1/g$,
\begin{equation} \label{Lofg}
\tL=\frac{1}{\ts-A}\left(g+a\ln g-a\ln(\ts-A)+b\right)\,,
\end{equation}
and insert it in (\ref{bL}). We obtain the final result
\begin{equation} \label{bfin}
\beta(g)=1-\frac{1}{g}\left(b+a(1+ \ln(\ts-A))-a\ln g\right)+{\cal
O}(1/g^2)\,.
\end{equation}
It is now obvious that the scaling function does indeed depend on the BCs
{\em via} the coefficients $a$ and $b$, and the dependence arises
at order $(\ln g)/g$.
Furthermore, $\beta(g)$ depends to order $1/g$ as well on the material
dependent dimensionless bulk conductivity $\ts$, and is therefore {\em
non--universal!} Again, the non--universality vanishes for $g\to\infty$
(equivalently, on the metallic side of the transition: $L\to\infty$), but is important if one is interested in
$\beta(g)$ at finite values of $g$.
Since HW BCs lead to smaller values of $\beta(g)$ at intermediate
values of $g$ than PBCs but to a smaller critical conductance, there should be a
point where the two curves cross, which would imply that in that point the
change of $g$ with the system size is independent of the BCs. Due to the dependence of $\beta(g)$ on $\ts$, this point is not expected to
be universal, though.

The most interesting question is of course, whether also the slope of
$\beta(g)$ at $g=g_c$ is changed by the BCs and/or $\ts$, as this slope
determines the critical
exponent $\nu$ defined by $\xi(w)\propto |g-g_c|^{-\nu}$ according to
$\beta(g)=\frac{1}{\nu}(g-g_c)/g_c$. This question
arises actually already from the dependence of spectral statistics on the
BCs, since the scaling function can be determined also from purely spectral
statistics \cite{Shklovskii93,Hofstetter93}. Very recently it has been
argued that within
the same universality class $\nu$ does at least not depend on the shape of
the sample \cite{Schreiber99}. Since  the critical
spectral statistics does depend on the shape of the sample much in the same
way as on the BCs \cite{Schweitzer98} (indeed, all that has been said above
about the dependence on the BCs translates one to one to a dependence on the
shape of the sample), one might suspect that $\nu$ is also
independent of the
BCs. On the other hand, considering the qualitative behavior of the two
scaling curves
a critical exponent independent of
the BCs would appear rather as coincidence.
However, so far it is an open question and definitly deserves attention
\cite{Keith00}.

With the dependence of the critical conductance distribution on the BCs, an
experimental test of the correctness of the one--parameter scaling picture
seems within reach. Even though an accurate absolute measurement of the
critical
exponent is rather difficult \cite{Thomas86,Katsumoto89}, one might hope to
detect a
{\em change} with the BCs. To this end it is not even necessary to
open and close the sample. Rather one can investigate the difference between
periodic and {\em anti--periodic} boundary conditions. At least in one
direction anti--periodic BCs, i.e.~a phase
factor $-1$ between two opposite sides of the sample, can be easily produced
by closing the sample
to a ring and introducing half a magentic flux quantum ($\phi=1/2$ in
(\ref{HAnd1})).
Note that for $\phi=1/2$ the system
still belongs  to the orthogonal universality class, since the Hamiltonian
has a real representation. This situation has been termed ``false time
reversal symmetry breaking'' \cite{Berry86}. An experimental search of a
change of the
scaling function in the metallic regime upon inclusion of half a flux
quantum would be as well a most welcome contribution to the long-lasting
debate on the limits of validity of one--parameter scaling.

In summary, we have shown that the conductance distribution at the Anderson
Metal--Insulator transition depends on the boundary conditions applied in
the directions transverse to the transport. Furthermore, in the metallic
regime the dependence of a change of the conductance with the system size
does not depend solely on the conductance itself but as well on the boundary
conditions and the dimensionless bulk conductivity. As a consequence the
scaling function $\beta(g)$ that describes the change of conductance when
the size of the sample is changed is not entirely universal but depends on
the boundary conditions and the amount of disorder in the sample from the
metallic regime up to the metal insulator transition.

{\em Acknowledgements:} D.B.~gratefully acknowledges useful discussions with
B.L.~Altshuler, P.A.~Braun and Y.V.~Fyodorov, and G.M.~would like to thank B.~Shapiro,
A.~Kamenev, V.~Kravtsov, and V.~Yudson. This work was supported by the
Sonderforschungsbereich 237 ``Unordnung und gro{\ss}e
Fluktuationen''. Numerical calculations were performed partly at the John
von Neumann institute for supercomputing in J\"ulich  and at the computer center IDRIS at Orsay (France).

\end{multicols}

\begin{table}[h]
\begin{center}
\begin{tabular}{c||c|c|}
&a&b\\\hline
PBC&$-5/(2\pi)\simeq-0.7958$&$5(\ln 3)/(4\pi)-2\cdot6.1509\pi^2\simeq -0.8093$\\\hline
HW&$1/(2\pi)\simeq 0.1592$&$ 2\cdot2.328/\pi^2-(\ln
3)/(4\pi)\simeq 0.3843$
\end{tabular}
\caption{Coefficients in the $1/\tL$ expansion of $g$ for periodic boundary
conditions (PBC) and hard walls (HW).\label{tabc}}
\end{center}
\end{table}


\begin{references}
\bibitem{Anderson58} P.W. Anderson, Phys.Rev.{\bf 109}, 1492 (1958).
\bibitem{Kramer93} For a review see B. Kramer and A. MacKinnon,
Rep. Prog. Phys.{\bf 56} 1469 (1993); Y. Imry in {\em mesoscopic
quantum physics}, Les Houches session LXI 1994, eds. E. Akkermans,
G. Montambaux, J.-L. Pichard and J. Zinn-Justin; North-Holland, Amsterdam
(1995); B.L. Altshuler and B.D. Simons, {\em ibid.}.
\bibitem{Abrahams79} E. Abrahams, P.W. Anderson, D.C. Licciardello, and
T.V. Ramakrishnan, Phys. Rev. Lett.{\bf 42}, 673 (1979); L.P. Gorkov,
A.I. Larkin, D.E. Khmelnitskii, JETP Lett.{\bf 30}, 228 (1979).
\bibitem{Wegner85} F.J. Wegner, in {\em Localisation, Interaction and Transport
Phenomena} (Springer Series in Solid State Science {\bf 61}) ed.~B. Kramer,
G. Bergmann, and Y. Bruynseraede, Springer, Berlin (1985).
\bibitem{Vollhardt82} D. Vollhardt and P. W\"olfle, Phys. Rev. Lett.{\bf 48}
699 (1982).
\bibitem{Shapiro86} B. Shapiro, Phys.Rev.B {\bf 34}, 4394 (1986); B. Shapiro,
Mod. Phys. Lett. B{\bf 5}, 687 (1991).
\bibitem{Markos94} P. Marko\v{s}, Europhys.Lett.{\bf 26}, 431 (1994).
\bibitem{Slevin97} K. Slevin and T. Ohtsuki, Phys. Rev. Lett.{\bf 78}, 4083
(1997).
\bibitem{Soukoulis99} C.M. Soukoulis, X. Wang, Q. Li, and M.M. Sigalas,
Phys.Rev.Lett.{\bf 82}, 668 (1999).
\bibitem{Slevin99} K. Slevin and  T. Ohtsuki, Phys. Rev. Lett. {\bf 82}, 669 (1999), Phys. Rev. Lett. {\bf 84}, 3915 (2000)
\bibitem{macKinnon85} A. MacKinnon, Z. Phys. {\bf B 59}, 385 (1985).
\bibitem{Braun98} D. Braun, G. Montambaux, and M. Pascaud, Phys. Rev. Lett.{\bf
81}, 1062 (1998).
\bibitem{Berry86} M.V. Berry and M. Robnick, J.Phys.A {\bf 19}, 649 (1986).
\bibitem{MacKinnon83} A. MacKinnon and B. Kramer, Z.Phys.B{\bf 53}, 1 (1983)
\bibitem{Braun97} D. Braun, E. Hofstetter, G. Montambaux, A. MacKinnon,
Phys. Rev. B{\bf 55}, 7557 (1997)
\bibitem{Shklovskii93} B.I. Shklovskii, B. Shapiro, B.R. Sears,
P. Lambrianides,
and H.B. Shore, Phys.Rev.B{\bf 47}, 11487 (1993).
\bibitem{Hofstetter93} E. Hofstetter and M. Schreiber, Phys.Rev.B {\bf 48},
16979 (1993).
\bibitem{Altshuler94} B.L. Altshuler and B.D. Simons, in {\em mesoscopic
quantum physics}, Les Houches session LXI 1994, eds. E. Akkermans,
G. Montambaux, J.-L. Pichard and J. Zinn-Justin; North-Holland, Amsterdam
(1995).
\bibitem{Kravtsov94} V.E. Kravtsov and A.D. Mirlin, JETP Lett. {\bf 60}, 656
(1994).
\bibitem{RG} I.S. Gradshteyn, I.M. Ryzhik, {\em Table of Integrals, Series,
and Products}, Fifth Edition, A. Jeffrey, ed., Academic Press, London (1996).
\bibitem{Kawabata85} A. Kawabata, Prog. Theor. Phys. Suppl.{\bf 84}, 16 (1985).
\bibitem{Lee85} P.A. Lee and T.V. Ramakrishnan, Rev. Mod. Phys., {\bf 57}, 287
(1985).
\bibitem{Imry94} Y. Imry in {\em mesoscopic quantum physics}, Les Houches
Session LXI, ed. E. Akkermans, G. Montambaux, J.--L. Pichard, and
J.Zinn--Justin, North--Holland, Amsterdam (1994).
\bibitem{Vollhardt92} D. Vollhardt and P. W\"olfle in {\em Electronic Phase
Transitions}, ed. W. Hanke and Ya.V. Kopaev, North--Holland, Amsterdam (1992).
\bibitem{Chandrasekhar91} V.Chandrasekhar, R.A. Webb, M.J. Brady, M.B. Ketchen,
W.J. Gallagher, and A. Kleinasser, Phys. Rev. Lett.{\bf 67}, 3578 (1991).
\bibitem{Keith00} While we were completing this manuscript, K. Slevin,
T. Ohtsuki and T. Kawarabayshi (Phys. Rev. Lett. {\bf 84}, 3915 (2000))
reported numerical evidence that the critical exponent does
{\em not} depend  on the BCs.
\bibitem{Thomas86} G.A. Thomas, {\em Localisation and Interactions in
Disordered and Doped Semiconductors}, ed. D.M. Finlayson, Edinburgh: SUSSP
(1986)
\bibitem{Katsumoto89} S. Katsumoto, F. Komori, N. Sano and S. Kobayashi,
J. Phys. Soc. Japan {\bf 56}, 2259 (1989).

\bibitem{Schreiber99}  F. Milde, R. A. Romer, M. Schreiber, V. Uski,
unpublished (cond-mat/9911029).
\bibitem{Schweitzer98}  H. Potempa, L. Schweitzer, J. Phys.: Condens. Matter
10, L431 (1998).
\end{references}
\end{document}